\documentclass[10pt,twocolumn,pre,aps,superscriptaddress]{revtex4-2}

\usepackage{amsmath}
\usepackage{graphicx}
\usepackage{color}
\usepackage{upgreek}
\usepackage{enumerate}
\usepackage[linkcolor = blue, citecolor = blue, urlcolor = blue, colorlinks = true]{hyperref}
\usepackage[version=3]{mhchem}
\usepackage{commath,amssymb}
\usepackage{ushort}
\usepackage{multirow}
\usepackage[usenames,dvipsnames]{xcolor}
\usepackage{cleveref}
\usepackage{epstopdf}
\usepackage{wasysym}
\usepackage[exponent-product = \cdot]{siunitx}

\begin{document}

\author{Joost de Graaf}
\email{j.degraaf@uu.nl}
\affiliation{Institute for Theoretical Physics, Center for Extreme Matter and Emergent Phenomena, Utrecht University, Princetonplein 5, 3584 CC Utrecht, The Netherlands}

\author{Kim William Torre}
\affiliation{Institute for Theoretical Physics, Center for Extreme Matter and Emergent Phenomena, Utrecht University, Princetonplein 5, 3584 CC Utrecht, The Netherlands}

\author{Wilson C.K. Poon}
\affiliation{SUPA, School of Physics and Astronomy, The University of Edinburgh, King's Buildings, Peter Guthrie Tait Road, Edinburgh, EH9 3FD, United Kingdom}

\author{Michiel Hermes}
\affiliation{Soft Condensed Matter, Debye Institute for Nanomaterials Science, Utrecht University, Princetonplein 1, 3584 CC Utrecht, The Netherlands}

\title{Hydrodynamic Stability Criterion for Colloidal Gelation under Gravity}

\date{\today}

\begin{abstract}
Attractive colloids diffuse and aggregate to form gels, solid-like particle networks suspended in a fluid. Gravity is known to strongly impact the stability of gels once they are formed. However, its effect on the process of gel formation has seldom been studied. Here, we simulate the effect of gravity on gelation using both Brownian dynamics and a lattice-Boltzmann algorithm that accounts for hydrodynamic interactions. We work in a confined geometry to capture macroscopic, buoyancy-induced flows driven by the density mismatch between fluid and colloids. These flows give rise to a stability criterion for network formation, based on an effective accelerated sedimentation of nascent clusters at low volume fractions that disrupts gelation. Above a critical volume fraction, mechanical strength in the forming gel network dominates the dynamics: the interface between the colloid-rich and colloid-poor region moves downward at an ever decreasing rate. Finally, we analyze the asymptotic state, the colloidal gel-like sediment, which we find not to be appreciably impacted by the vigorous flows that can occur during the settling of the colloids. Our findings represent the first steps toward understanding how flow during formation affects the life span of colloidal gels.
\end{abstract}

\maketitle

\section{\label{sec:intro}Introduction}

Colloidal gels feature in a range of applications, because their properties bridge fluid and solid response. The arrested dynamics in the percolating particle network allow a gel to support its own weight for a finite time (the `shelf life'), while its mechanical weakness allows it to be poured upon the application of relatively low stresses. A large body of experimental, simulation, and theoretical work has elucidated colloidal gel formation~\cite{allain1993effects, poon1995gelation, Verhaegh1997, manley2005glasslike, zaccarelli2007colloidal, Lu2008gelation, laurati2009structure, zhang2019correlated, wang2019surface, immink2020using, nguyen2020computer}, rheology~\cite{kamp2009universal, laurati2011nonlinear, koumakis2015tuning, moghimi2017colloidal, diba2017highly, vandoorn2018strand, verweij2019plasticity}, and shelf life~\cite{starrs2002collapse, manley2005gravitational, kim2007gravitational, bailey2007spinodal, buscall2009towards, brambilla2011highly, bartlett2012sudden, teece2014gels, secchi2014time, harich2016gravitational, varga2018large, varga2018modelling, padmanabhan2018gravitational, tsurusawa2019direct, sui2020dynamic}. Many of these studies point toward fluid flow and hydrodynamic interactions (HIs) as contributing to or even dominating the properties and dynamics of colloidal gels.

Until recently, however, simulations of colloidal gelation have mostly neglected HIs and flow, due to computational limitations. State-of-the-art computational fluid dynamics methods, capable of simulating many thousands of suspended colloids, have started to rectify this~\cite{royall2015probing, varga2015hydrodynamics, varga2018modelling, degraaf2019hydrodynamics, turetta2022role, alcazar2022hydrodynamics}. For example, our lattice-Boltzmann (LB) simulations~\cite{degraaf2019hydrodynamics} have shown that HIs speed up the gelation of particles with strong, short-ranged attractions at low colloid volume fractions $\phi_{0}$ and slow it down for high $\phi_{0}$, with respect to systems without such interactions. Surprisingly for an out-of-equilibrium process, the network structure with and without HIs appeared identical when the systems are compared at equal `structural time'. These findings likely resolve seemingly conflicting results~\cite{yamamoto2008role, furukawa2010key, varga2015hydrodynamics} on the role of HIs in establishing structure during colloidal gelation.

In practice, hydrodynamic flows are almost always primarily driven by buoyancy forces $F_{\mathrm{B}}$. That is, there is a mass density difference $\Delta \rho = \rho_{\mathrm{c}} - \rho_{s}$ between the colloids ($\rho_{\mathrm{c}}$) and the suspending medium ($\rho_{s}$). For spherical colloids of diameter $\sigma$, this leads to 
\begin{align}
\label{eq:Fbuoy} F_{\mathrm{B}} &= \frac{\pi}{6} g \Delta \rho \sigma^{3} < 0 ,
\end{align}
where $g$ is the local gravitational constant. It is already known that buoyancy-driven flows are implicated in the way gels fail at the end of their life time. Such collapse can involve recirculation of the suspending medium triggered by the falling of dense `debris' from the air-suspension meniscus~\cite{starrs2002collapse, bartlett2012sudden, secchi2014time, harich2016gravitational}. The question naturally arises: ``How do such flows affect the process by which gels form in the first place?''

Allain~\textit{et~al.}~addressed this using a diffusion-limited-cluster-aggregation (DLCA) model~\cite{allain1995aggregation}, identifying a transition between ``cluster deposition'' and ``collective settling''. The transition was argued to occur when the gelation, sedimentation, and diffusion times are equal, leading to a crossover volume fraction
\begin{align}
\label{eq:phist} \phi^{\ast} &\propto \mathrm{Pe}^{(3-d)/(1+d)} , 
\end{align}
with $d$ the fractal dimension ($d = 1.8$ in DLCA). The gravitational P{\'e}clet number 
\begin{align}
\label{eq:Peg} \mathrm{Pe} &= \frac{ \pi g \Delta \rho \sigma^{4} }{12 k_{\mathrm{B}}T} ,
\end{align}
is the ratio between displacement through $F_{\mathrm{B}}$ and thermal diffusion with diffusivity $D = k_{\mathrm{B}}T / ( 3 \pi \eta \sigma )$, where $T$ is the temperature, $k_{\mathrm{B}}$ Boltzmann's constant, and $\eta$ the dynamic viscosity of the suspending medium. The theoretical prediction order-of-magnitude matched the experimental lower stability bound. The agreement is, however, of limited utility as the DLCA picture only applies in the limit $\phi_{0} \to 0$, where $\phi_{0}$ is understood to be the homogeneous colloid volume fraction at preparation.

The settling of aggregating colloids and the nature of the sediment has also been the subject of several numerical investigations~\cite{gonzalez2002colloidal, gonzalez2004colloidal, moncho2010effects, whitmer2011sedimentation, moncho2012peclet}, recently including HIs~\cite{fiore2018hindered, turetta2022role}. When attractions are too weak to gel the system, the \textit{steady-state} sedimentation rate can be slightly enhanced with respect to that of a single particle at low $\phi_{0}$~\cite{moncho2010effects, moncho2012peclet, fiore2018hindered}. For sufficiently strong attractions, HIs can even interrupt the formation of a percolating network by aligning and reconfiguring the forming clusters~\cite{turetta2022role}. However, the full process of gelation in a confining geometry with (emerging) height-dependent density and flow heterogeneities has not been studied thus far.

Here, we extend our recent computational analysis of bulk gelation of colloids with short-ranged, depletion-like attractions~\cite{degraaf2019hydrodynamics} to include buoyancy ($F_{\mathrm{B}} > 0$). We compared simulations using overdamped Langevin dynamics (no HIs or flow: NH;~\footnote{Our parameter choices make it effectively a Brownian dynamics simulation.}) to those using a GPU-accelerated fluctuating LB fluid (with HIs and flow: WH) that capture large-scale flow and hydrodynamic interactions between colloids. We specifically used an enclosed simulation volume to ensure that density heterogeneities throughout the sample are accounted for, which can impact the large-scale flows. These simulations allowed us to identify a distinct gelation criterion, given by a critical initial volume fraction $\phi_{\mathrm{c}}$, based on the (initial) \textit{non-steady} settling of the suspension.

Below $\phi_{\mathrm{c}}$, the settling is characterized by an effective acceleration of the colloid-rich phase toward the bottom of the sample. The buoyancy-driven flows are sufficiently vigorous to eject small clusters from the interface between the colloid-rich and colloid-poor phases and generally disrupt the formation of the gel network. The acceleration in a system that otherwise obeys Stokes-flow conditions, can be explained by the combined effect of cluster growth and cluster reorientation. We present a minimal theoretical model to help clarify this effect. Above $\phi_{\mathrm{c}}$, the interface settling is decelerated, which we associate with mechanical strength in the nascent gel dominating the dynamics. Surprisingly, despite the strong effect of fluid flow at low volume fractions, the colloidal sediment that is achieved for long times compared to the time it takes a single colloid to sediment its own diameter, is not appreciably affected. That is, this sediment is identical within the error with or without HIs and flow. It is this sediment that behaves as a true, albeit density-wise heterogeneous gel,~\textit{i.e.}, it has strongly arrested dynamics at intermediate volume fractions $\phi \approx 0.3$.

We close our analysis with a comparison to other simulation studies and by highlighting the limitations of our work. We also connect to recent experiments on gelation in small confining geometries, thus providing a foundation for further in-depth studies.

\section{\label{sec:method}Numerical Methods}

The simulation methods used here to study the effect of buoyancy on the formation of attractive colloidal gels are similar to those we employed recently~\cite{degraaf2019hydrodynamics}. We will therefore briefly outline the key points and focus on the differences that are needed to account for $F_{\mathrm{B}} > 0$. 

Our simulation volume is a prism with a square base of length $32 \sigma$ and it is periodic in the basal direction. We also performed a few larger size simulations with basal dimensions $48\sigma$ and $64\sigma$, respectively. This volume is enclosed vertically by a `floor' and `ceiling' that are purely repulsive to the colloids and are impenetrable to the fluid, possessing no-slip boundary conditions its velocity field (standard LB bounce-back algorithm). The placement of the enclosing boundaries is such that the colloid centers are constrained to a volume with height $H = 125\sigma$. Additionally, a few simulations were performed with effective heights $H = 189\sigma$ and $H = 253\sigma$, respectively. We will indicate it in the text when we deviate from our standard choices of $32 \sigma$ width and $H = 125\sigma$.

\begin{figure*}[!htb]
\centering
\includegraphics[width=1.9\columnwidth]{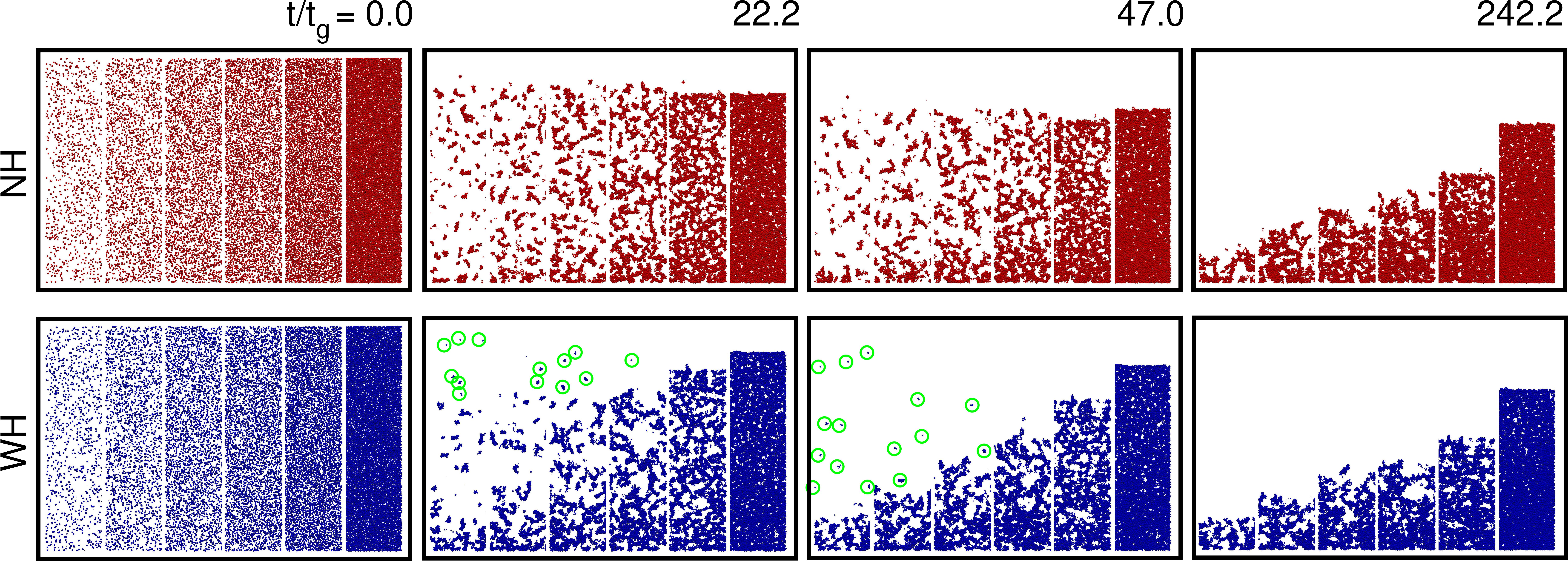}
\caption{\label{fig:collapse}Snapshots of the colloids in a thin (2$\sigma$-wide) slice though the center of the simulation box with the colloids indicated in red (NH; top) or blue (WH; bottom); $\mathrm{Pe} = 0.28$ and $\epsilon = 10 k_{\mathrm{B}}T$. In a single panel, the initial volume fraction of the sample increases from left to right: $\phi_{0} = 0.025$, $0.05$, $0.075$, $0.1$, $0.15$, and $0.3$, respectively, as indicated. Each time series has 4 panels, showing characteristic behaviors between the initial and final configuration: $t/t_{\mathrm{g}} = 0$, $22.2$, $47.0$, and $242.2$, respectively. For the hydrodynamic simulations small clusters and single particles are expelled from the forming gel (encircled in green).}
\end{figure*}

We choose an enclosed, rather than fully periodic simulation volume to facilitate the modeling of backflows throughout the sample. The fully periodic setup used by Varga, Hofmann, and Swan~\cite{varga2018modelling} to study gel rupture and more recently by Turetta and Lattuada~\cite{turetta2022role} to study colloidal gelation --- both using $F_{\mathrm{B}} > 0$ --- imposes global momentum conservation within the simulation volume. However, because a local segment of gel experiences flows that result from compaction throughout the sample, such a local conservation argument may not hold. That is, the system may not reach steady-state settling: (i) Heterogeneous flows can damage the nascent gel locally, setting up larger recirculatory flows driven by gravity. (ii) Typically settling gels have a heterogeneous density, meaning that flow through a specific slice of the sample can be driven by forces that originate lower (deeper) in the sample. Both of these considerations informed our decision \textit{not} to follow the zero-net-momentum approach here. We also focus on gel formation rather than rupture, as we found that there is significant impact on the amount of preforming of a gel in our (unreported) attempts to reproduce the results of Ref.~\cite{varga2018modelling}. We will return to the topic of preforming gels later.

We match the bulk diffusivity of our particles between the NH and WH simulations by setting the viscosity of our suspending medium $\eta$, such that $D = k_{\mathrm{B}}T / ( 3 \pi \eta \sigma ) = 1.1\times10^{-2} \sigma^{2} \tau^{-1}$, with the thermal energy $k_{\mathrm{B}}T \equiv 1$, unit time $\tau \equiv \sqrt{m \sigma^{2} / (k_{\mathrm{B}}T)}$, and particle mass $m = (\pi/6) \rho_{\mathrm{c}} \sigma^{3} \equiv 1$. A short-ranged attraction between the colloids is modeled using~\cite{degraaf2019hydrodynamics}:
\begin{align}
\label{eq:LJgen} U_{\mathrm{LJ}}^{\mathrm{gen}}(r) &=
\begin{cases}
\displaystyle \epsilon \left[ \left( \frac{\sigma}{r} \right)^{96} - 2 \left( \frac{\sigma}{r} \right)^{48} + c \right] & r < r_{\mathrm{c}} \\
0 & r \ge r_{\mathrm{c}} 
\end{cases} ,
\end{align}
which mimics an Asakura-Oosawa interaction~\cite{asakura1958interaction}. Here, $r$ is the separation, $r_{\mathrm{c}}$ the cut-off distance, $\epsilon$ the attraction strength, and $c$ a shift. The confining boundaries interact with the colloids \textit{via} Eq.~\eqref{eq:LJgen}, except that $r$ has been replaced by the center-to-boundary distance and we use repulsive parameters $\epsilon = 10 k_{\mathrm{B}}T$, $c = 1$, and $r_{\mathrm{c}} = \sigma$.

The choice for purely repulsive boundaries on the top and bottom wall of the simulation volume avoids the additional complication of the gel `hanging off' the ceiling. Padmanabhan and Zia~\cite{padmanabhan2018gravitational} reported this to affect the dynamics in non-hydrodynamic simulations. We verified that our results are not strongly impacted by neglecting wall attraction, by performing additional simulations where there was a $10 k_{\mathrm{B}}T$ and $20 k_{\mathrm{B}}T$ colloid-wall attraction; both for a colloid-colloid attraction of $\epsilon = 10 k_{\mathrm{B}}T$. The shape of the colloid-wall potential is that of our high-exponent, shifted Lennard-Jones interaction provided in Eq.~\eqref{eq:LJgen}; replacing $r$ with the distance between the colloid and wall. In these attractive-wall simulations, we found that for a sufficiently high colloid-wall attraction, clusters were left behind on the top wall as the gel settled. However, this top-wall deposition did not result in an appreciable difference in the settling, which is in line with our finding that the nascent gels are not able to support their own weight. In view of this, we report only on purely repulsive walls in the remainder of this paper.

Gravity is introduced by imposing a constant force on the colloids in the negative $z$-direction: $-F_{\mathrm{B}} \hat{z}$. The associated time scale is the gravitational time, the time it takes a particle to sediment its own radius $t_{\mathrm{g}} = 3 \pi \eta \sigma / \vert F_{\rm B} \vert$. We use a sufficiently small $\vert F_{\rm B} \vert$ to stay in the low-Reynolds-number regime ($\mathrm{Re} \lesssim 0.05$ throughout),~\textit{i.e.}, hydrodynamic dissipation dominates inertia.

We consider non-equidistant initial $\phi_{0} \in [0.01,0.3]$ corresponding to between 2,503 and 75,098 colloids in our simulation volume, and four values of $\mathrm{Pe} = 0.03$, $0.06$, $0.28$, and $1.42$, respectively, with the last two chosen to correspond to recent experiments~\cite{zhou2018onset,harich2016gravitational}. Our systems are equilibrated using a repulsive potential: $\epsilon = 10 k_{\mathrm{B}}T$, $c = 1$, and $r_{\mathrm{c}} = \sigma$, for $50 t_{\mathrm{B}}$ before quenching the system. Here, $t_{\mathrm{B}} = \sigma^{2}/(4 D) = 23.5 \tau$ is the (Brownian) time for a particle to diffuse its own radius. At a quench, we simultaneously switch on gravity and an attraction of $\epsilon = 10 k_{\mathrm{B}}T$ (or $\epsilon = 20 k_{\mathrm{B}}T$); $c = 0$, and $r_{\mathrm{c}} = 1.5\sigma$. In all cases, we performed 5 independent simulations.

Our runs typically took several hours to several days to run on a desktop (i7-8700) with modern GPU (NVidia RTX 2080 Ti). Smaller values of the gravitational P{\'e}clet number $\mathrm{Pe}$ and larger values of the number of particles $N$ lead to longer simulation times, as expected. The increases are such that our smallest values of $\mathrm{Pe}$ are at the edge of what is computationally feasible for the largest number of particles $N$. The accompanying scripts in the data package provide the means to reproduce our results.

\section{\label{sec:buoyant}Buoyant Settling of Gels}

Figure~\ref{fig:collapse} shows the typical behavior with and without HIs over a range of $\phi_{0}$ (videos are in the supplement~\cite{supplement} and described in Appendix~\ref{sec:videos}). The interface between the colloid-rich and colloid-poor regions settles at roughly the same rate independent of $\phi_{0}$ in systems without flow. Apparent differences between the samples at low $\phi_{0}$ can be attributed to a diffusion-based widening of the interface during settling, as the network structure has not yet become fully arrested. With flow, the situation changes drastically. The systems with the lowest $\phi_{0}$ sediment fastest and fluid backflow causes clusters to be expelled from the colloid-rich region.

\begin{figure}[!htb]
\centering
\includegraphics[width=0.95\columnwidth]{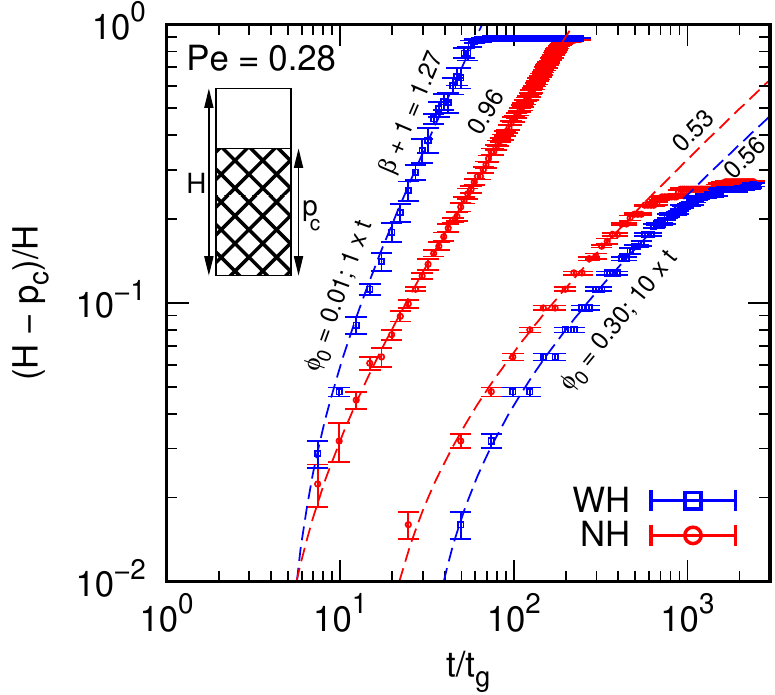}
\caption{\label{fig:interface}The effect of buoyancy on the interface between the colloid-rich and colloid-poor region for $\mathrm{Pe} = 0.28$ and $\epsilon = 10 k_{\mathrm{B}}T$. The reduced height of the colloid-poor region $(H - p_{\mathrm{c}})/H$ is given as a function of the reduced time $t/t_{\mathrm{g}}$ for two initial colloid volume fractions $\phi_{0} = 0.01$ and $\phi_{0} = 0.3$, as labeled. For $\phi_{0} = 0.3$, time is multiplied by a factor $10$ to aid the presentation. The dashed lines indicate shifted power-law fits, see main text. The mean fitted power-law coefficient $\beta + 1$ is given by the numbers next to each line. The inset defines the height $H$ and interface position $p_{\mathrm{c}}$.}
\end{figure}

We quantified this difference as follows. The data is binned and averaged over slices $4\sigma$ in height --- accounting for near-wall density reductions due to excluded volume --- to obtain the height profile $\phi(z)$. We consider slices with $\phi(z) > \phi_{0}/2$ as colloid rich and determine the position of the interface $p_{\mathrm{c}}$ by extracting highest colloid-rich bin, averaging this over the 5 runs. The interface position at small $t$ is well described by $p_{\mathrm{c}} = H - A - Bt^{\beta + 1}$, where $H$ is the initial height and $A, B$ are constants, see Fig.~\ref{fig:interface}. Appendix~\ref{sec:scale} provides evidence that this initial settling of the interface is not dependent on $H$.

\section{\label{sec:critical}Critical Volume Fraction}

\begin{figure}[!htb]
\centering
\includegraphics[width=0.95\columnwidth]{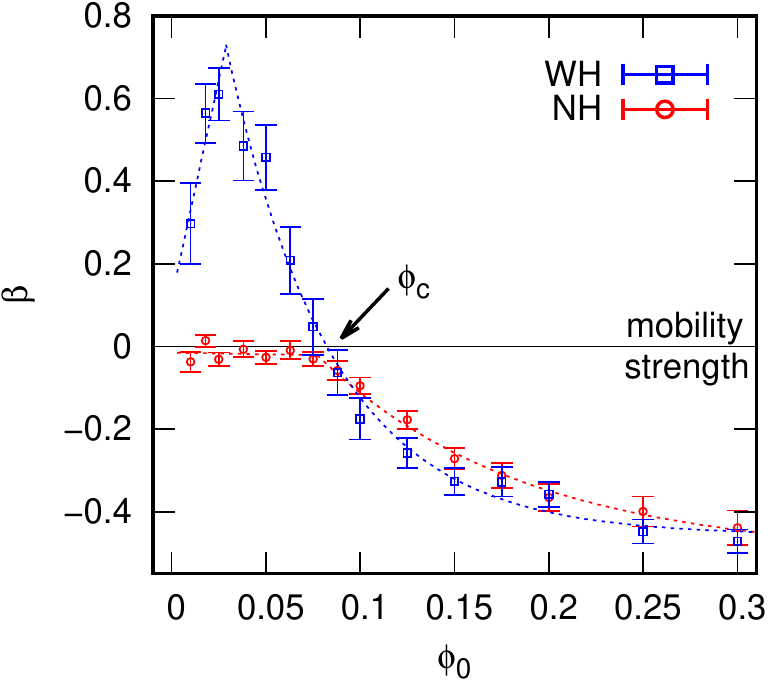}
\caption{\label{fig:crit} The exponent $\beta$ for $\mathrm{Pe} = 0.28$ and $\epsilon = 10 k_{\mathrm{B}}T$ as a function of $\phi_{0}$ with (\textcolor{blue}{$\Square$}) and without (\textcolor{red}{$\Circle$}) HIs, showing standard errors. Dashed curves: piece-wise fits as guides to the eye. Arrow: critical crossover volume fraction $\phi_{\mathrm{c}}$.}
\end{figure}

The form of $p_{\mathrm{c}}(t)$ implies that the interface velocity $v_{\mathrm{c}}$ scales as $t^{\beta}$. Interestingly, $\beta(\phi_{0})$ is non-monotonic, see Fig.~\ref{fig:crit}. Unsurprisingly, $\beta \approx 0$ for the NH system at low $\phi_{0}$: the interface simply tracks the sedimentation of individual colloids and $v_{\mathrm{c}} \lesssim \sigma / (2 t_{\mathrm{g}})$. With HIs and fluid flow, however, $\beta > 0$ for systems at low initial volume fractions, indicating accelerating sedimentation. We will return to this in Section~\ref{sec:theory}.

At some critical $\phi_{\mathrm{c}}$, $\beta$ vanishes and subsequently becomes negative, decreasing further with increasing $\phi_{0}$. This decrease appears to level off and the WH and NH systems follow nearly the same trend within the error. Note that the WH system appears to sediment slightly slower (equal $\beta$, smaller $v_{c}$). This is probably due to squeeze flows, which hinder the colloids from approaching each other closely~\cite{degraaf2019hydrodynamics}. It should be noted that our lattice-Boltzmann variant does not resolve lubrication flows, but an effective increase of the friction upon approach is nonetheless present. Negative $\beta$ indicates decelerating sedimentation, which is a necessary (but not sufficient) condition for gelation. In all cases, $p_{\mathrm{c}}$ decreases without showing any plateau or inflection point that can be linked to a (short-lived) arrested state.

We have fitted a linear function and an exponential decay to $\beta(\phi_{0})$ using least squares in the low- and high-$\phi_{0}$ regions, respectively, see Fig.~\ref{fig:crit}. The former is motivated by the more pronounced linear trend observed for $\mathrm{Pe} = 1.42$, see Appendix~\ref{sec:exponents}, which provides $\beta(\phi_{0})$ curves for different values of $\mathrm{Pe}$. 

\section{\label{sec:theory}Accelerated Sedimentation}

Before moving onto our dynamic stability criterion, let us consider the accelerated nature of the settling of the interface between the colloid-rich and colloid-poor region in our suspension for low $\phi_{0}$. We will argue using a minimal theoretical model that this is indicative of \textit{anisotropic} collision-based settling and aggregation in the direction of gravity. In Section~\ref{sec:clusters}, we confront this understanding with our simulation data.

Note that acceleration at low Reynolds numbers ($\mathrm{Re} \lesssim 0.05$) is counterintuitive. In this regime, inertia is dominated completely by friction: the fluid and particles suspended therein do not accelerate. However, other simulation studies have shown that for weak attractions between the colloids, the \textit{steady-state} sedimentation rate can be slightly enhanced with respect to that of a single particle at low $\phi_{0}$~\cite{moncho2010effects, moncho2012peclet, fiore2018hindered}. Clearly, there should be a transient regime upon quenching a system between reduced settling of `free' particles and that of the settling steady-state aggregates.

In the case of sufficiently strong interactions to induce gelation, Turetta and Lattuada~\cite{turetta2022role} indicate the emergence of a steady state with an enhanced settling rate. In this state, hydrodynamic interactions impact shape and orientation of the clusters, as well as their ability to percolate. However, the acceleration of the interface itself is thus far unaddressed and it is necessary to explain the absence of a steady-state settling behavior in our system. We do so by considering an idealized one-dimensional (1D) argument.

Assume that $k$ spheres in contact form a `rod' of length $l = 2 k a$, which is aligned in the direction of gravity ($k$ measures the rod's aspect ratio). This rod has a longitudinal friction coefficient of 
\begin{equation}
\label{eq:rod} \gamma_{\parallel} \approx \frac{3}{2} \gamma_{\mathrm{sph}} \frac{k}{\log k},
\end{equation}
in bulk fluid~\cite{doi1986theory}. Here, $\gamma_{\mathrm{sph}} = 6 \pi \eta a$ is the single-sphere sedimentation coefficient and $\log$ represents the natural logarithm. Note that the net buoyant force also scales with $k$, which implies that the sedimentation velocity of a single rod is $\propto \log k$.

Now we assume that this infinite dilution result also holds for slightly higher densities and we consider several vertically aligned rods with a distributions of lengths. In particular, we assume that Eq.~\eqref{eq:rod} holds for densities that are sufficient to enable accumulation by collision. That is, longer rods sediment faster, accumulating shorter rods sedimenting at a lower level as they overtake these. This increases their length, since the accumulation is in the length-wise direction, and consequently their sedimentation rate. This in turn increases the rate of accumulation, leading to an acceleration of settling.

In our idealized model, the process only terminates if the initial state permits a steady-state distribution, wherein rods of equal length sediment with fixed separations. Note that hydrodynamic interactions could, in practice, destabilize such a configuration even if all rods have fixed length, but are not equidistantly spaced. In our much more involved simulations, this situation cannot occur. The nascent gel strands are deformed, reorient, break due to hydrodynamic interaction and backflow. But, more importantly, the system densifies leading to the formation of a network that can support itself. Both aspects would prevent the formation of a steady state of settling, which is why we chose to perform our analysis in a confining geometry, also see Section~\ref{sec:method}.

\section{\label{sec:clusters}Early-Time Clustering}

\begin{figure}[!h]
\centering
\includegraphics[width=0.95\columnwidth]{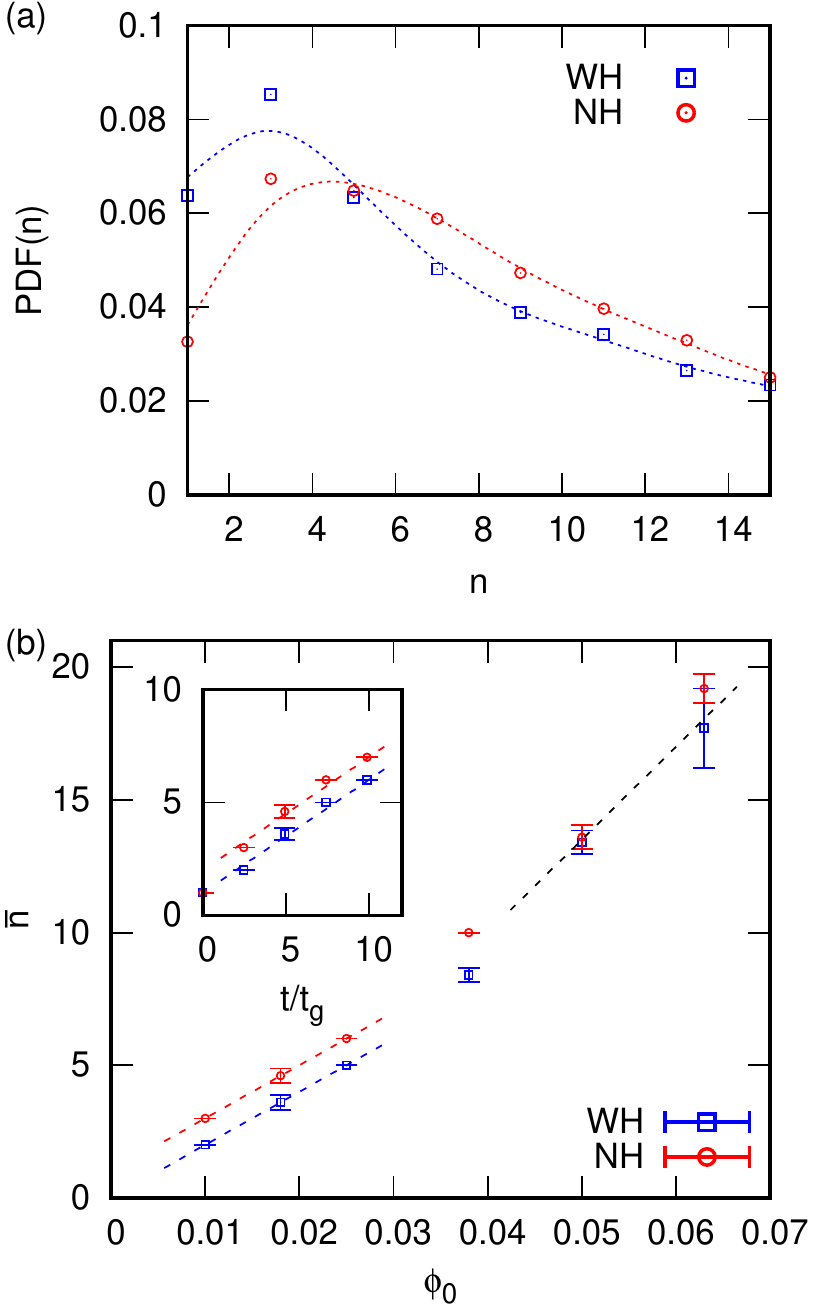}
\caption{\label{fig:clust}Properties of the clusters that form close to time $t = 0$ for a sample with $\mathrm{Pe} = 0.28$. In all cases, blue squares indicate data obtained with hydrodynamics (WH) and red circles the non-hydrodynamic (NH) data. (a) The probability density function (PDF) for free clusters of size $n$ at time $t \approx 5 t_{\mathrm{g}}$ for colloid volume fraction $\phi_{0} = 0.05$. The dashed lines serve as guides to the eye. (b) The median value $\bar{n}$ of the PDF for the cluster size as a function of the initial volume fraction $\phi_{0}$ ($t = 5 t_{\mathrm{g}}$) and time $t$ (inset; $\phi_{0} = 0.018$) for $\mathrm{Pe} = 0.28$. The error bars provide the standard error of the mean and the dashed lines are linear fits.}
\end{figure}

We test the validity of our simple argument by studying the early-time clustering in our simulations. In the following, we use $\phi_{0} = 0.05$ and $\mathrm{Pe} = 0.28$ to ensure that we are in a regime of (maximally) accelerated settling. Figure~\ref{fig:clust}a shows a representative example of the difference in the free-cluster distribution between systems with and without HIs ($t \approx 5 t_{\mathrm{g}}$). The `free' clusters were extracted by tallying all clusters of size $n$ in the system and removing the large, single cluster that constitutes the sediment and parts of the gel that are forming near the bottom of the sample. To obtain accurate statistics, data from all 5 simulation runs were combined, which allowed us to arrive at the probability density function by normalizing the resulting histogram.

Note that the systems with HIs have smaller clusters, presumably due to backflows breaking up the nascent gel strands. This does not invalidate the argument we put forward in Section~\ref{sec:theory}, however, as the temporal evolution of the full distribution should be examined to judge the effect of HIs and flow. Additionally, the small clusters are typically expelled out of the interface by being carried along with the backflow. That is, these small clusters are a poor indicator of the net effect. Instead, we concentrate on the median cluster size $\bar{n}$ that can be extracted from the probability density functions (PDFs) that can be computed for each time $t$.

Figure~\ref{fig:clust}b shows an example of the behavior of $\bar{n}$. The main panel of this graph shows that $\bar{n} \propto \phi_{0}$ for low volume fractions, where there is a small difference in prefactor between systems with and without HIs. At slightly higher volume fractions the trend appears to be linear still, but within the error, the distinction between the two systems has blurred. In addition, there is a change in slope that is not relevant to the discussion to follow. The linearity $\bar{n} \propto \phi_{0}$ could be related to the constant settling predication proposed by Allain~\textit{et~al.}~\cite{allain1995aggregation}. However, note that their argument is based on \textit{spherical} aggregates, which we will shortly see is not appropriate for describing our system. In addition, as we will also see in Section~\ref{sec:stable}, our data does not follow the DLCA trend that Allain~\textit{et al.}~base their argument on. Thus, the difference must be sought in other quantifiers.

The inset to Fig.~\ref{fig:clust}b reveals that at short times $\bar{n} \propto t$ for small volume fractions ($\phi_{0} = 0.018$) both with and without flow and HIs. We conclude that in both cases the clusters grow and that at these short times, there appears to be no intrinsic difference between the mechanism of growth. The NH case has slightly faster cluster growth, again, presumably due to the disruptive effect of backflow interfering with growth in the WH case. This is not entirely unexpected, because for a value of $\mathrm{Pe} = 0.28$, diffusion outstrips sedimentation by a factor of four on the single-particle level.

For each free cluster, we determined the center (of mass) and computed the distance vectors $\boldsymbol{\Delta r}$ of all particles in the cluster with respect to this center. Next, we computed the cosine of the polar angle with respect to the $z$-axis (along which gravity is pointed): $\cos \theta = \boldsymbol{\hat{z}} \cdot \boldsymbol{\Delta r} / \vert \boldsymbol{\Delta r} \vert$, where $\boldsymbol{\hat{z}}$ is a unit vector. Finally, we averaged the value $P_{2}(\cos \theta)$, where $P_{2}$ is the second Legendre polynomial, over all clusters of the same size for a given time to arrive at the mean cluster orientation $\bar{p}_{2}$. This quantity is an indicator of the amount, by which the clusters are elongated along the $z$-axis. When $\bar{p}_{2} = 0$, there is no intrinsic bias, clusters that have $\bar{p}_{2} \approx -1/2$ are flattened out in the $xy$-plane, and clusters with $\bar{p}_{2} = 1$ are fully aligned with the $z$-axis.

\begin{figure}[t]
\centering
\includegraphics[width=0.95\columnwidth]{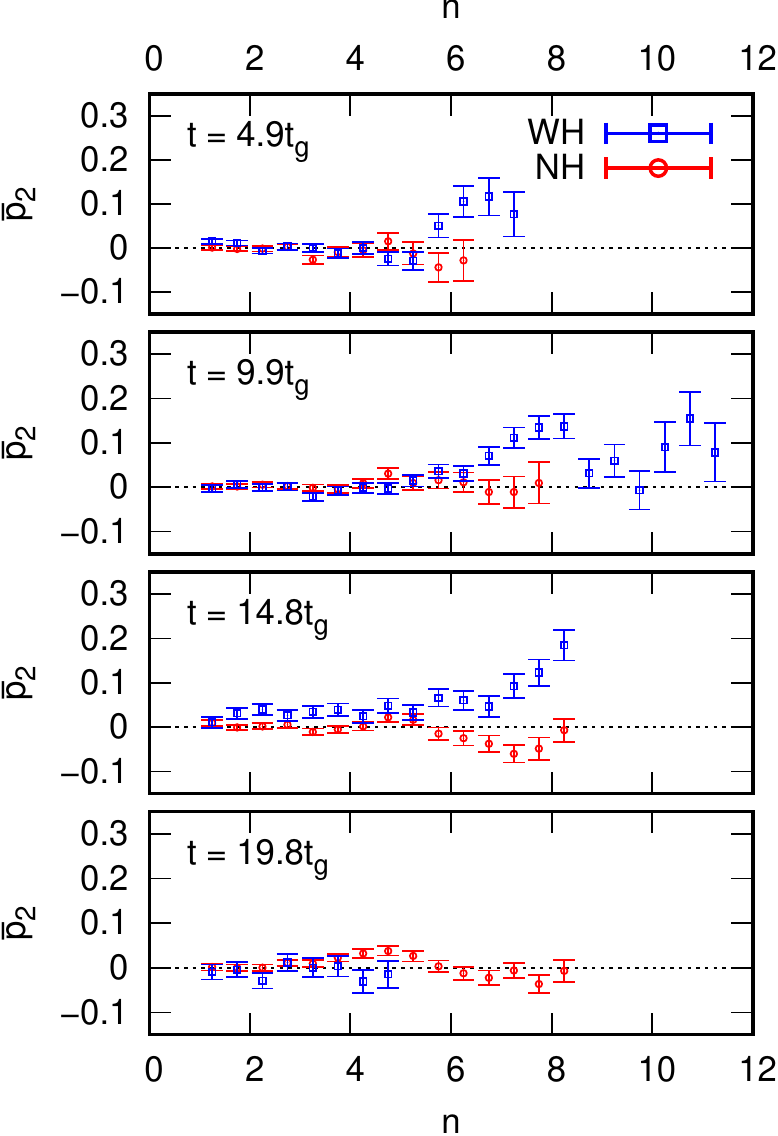}
\caption{\label{fig:orient}The average orientation $\bar{p}_{2}$ --- as defined in the main text --- of the `free' clusters in the sample as a function of time $t$ (expressed in gravitational times $t_{\mathrm{g}}$) and cluster size $n$. Simulations were performed with $\mathrm{Pe} = 0.28$ at $\phi_{0} = 0.05$, blue squares indicate data obtained with hydrodynamics (WH) and red circles the non-hydrodynamic (NH) data. The dashed black line shows the mean $\bar{p}_{2}$ value for an isotropic cluster and the error bars provide the standard error of the mean.}
\end{figure}

Figure~\ref{fig:orient} shows the value of $\bar{p}_{2}$ for several times $t$ post quenching the system. In simulations without HIs and flow, we find that there is no bias to the cluster orientation within the error for all times. The data obtained from our hydrodynamic simulations reveals that small clusters (up to about $n=5$ in size) have no intrinsic bias in their orientation. However, for times $t \lesssim 17.5 t_{\mathrm{g}}$, when most of the colloidal clusters have not yet fully settled, there is a noticeable difference between the simulations with and without HIs. Around $t=5t_{\mathrm{g}}$, we find slightly larger clusters in the system with HIs. These continue to grow and, around $t = 10t_{\mathrm{g}}$, display a measurable anisotropy, favoring orientation along the $z$-axis. The largest clusters reach the bottom sediment before $t \approx 15 t_{\mathrm{g}}$, but those larger clusters that remain suspended in the simulation with HIs, show an increased alignment with gravity. At $t \approx 20 t_{\mathrm{g}}$, most of the large clusters in the simulations with HIs have sedimented and there no longer is an appreciable bias.

Combining the understanding that our minimal 1D model brings with the cluster properties that we have shown here, we conclude that flow and hydrodynamic interactions, potentially coupled with mobility-mediated growth, indeed lead to an orientational bias in sufficiently large clusters. As predicted by our 1D model, the oriented clusters settle out rapidly, which underlies the accelerated sedimentation of the interface between the colloid-rich and colloid-poor parts of the suspension. Our result can be seen as a transient variant of the observations by Turetta and Lattuada~\cite{turetta2022role}.

\section{\label{sec:stable}Dynamical Stability Criterion}

\begin{figure}[t]
\centering
\includegraphics[width=0.95\columnwidth]{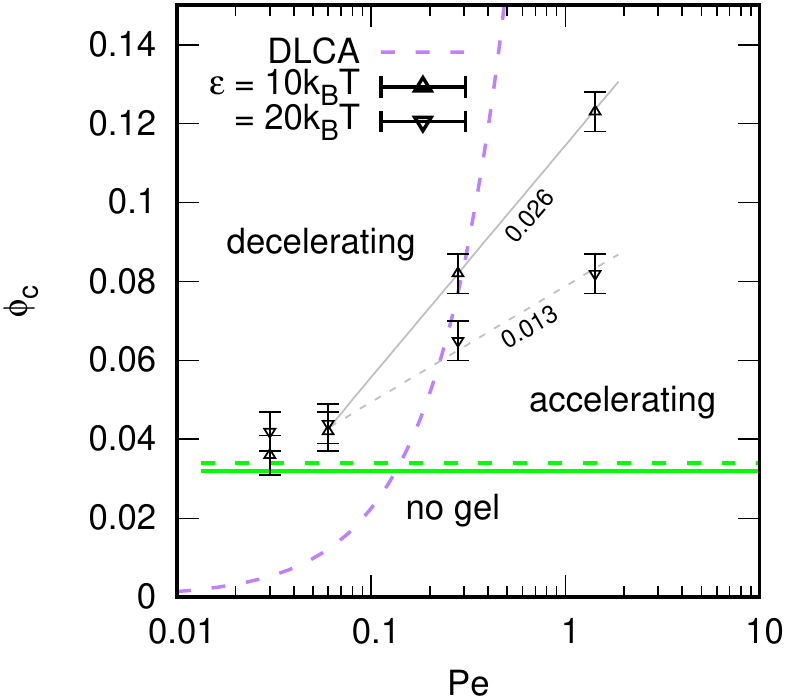}
\caption{\label{fig:cross}The critical volume fraction $\phi_{\mathrm{c}}$ as a function $\mathrm{Pe}$ for $\epsilon = 10 k_{\mathrm{B}}T$ ($\triangle$) and $20 k_{\mathrm{B}}T$ ($\triangledown$) showing standard errors. Fitted lines (grey): $\phi_{\mathrm{c}} \propto \log \mathrm{Pe}$ labelled by the prefactor. Green line: the bulk percolation threshold, $\phi = 0.032$ and $0.034$ for $\epsilon = 10$ and $20 k_{\mathrm{B}}T$, respectively. Purple dashed curve: DLCA prediction~\cite{allain1995aggregation} with unit prefactor and $d = 1.8$.}
\end{figure}

Returning to the dynamics of the interface, we can use the specific value of $\phi_{0}$ for which $\beta = 0$ to define the crossover volume fraction $\phi_{\mathrm{c}}$. We determine this value from our exponential fit, see Fig.~\ref{fig:crit}. By doing so for a range of P{\'e}clet numbers and two values of $\epsilon$, we are able to create a state diagram for the dynamical stability of a colloidal gel, see Fig.~\ref{fig:cross}.

We find a regime where $\phi_{\mathrm{c}} \propto \log \mathrm{Pe}$, so that a straight line demarcates an `accelerating' from a `decelerating' settling region in the $(\log \mathrm{Pe},\phi_{\mathrm{c}})$ state space of gelation/sedimentation. In the `accelerating' region, particle networks can form in principle, but the nascent structures are destroyed in practice by buoyancy-induced back flows, as evidenced by small clusters being expelled from the gel, see Fig.~\ref{fig:collapse} and supplemental movies~\cite{supplement}. In the `decelerating' region, networks can form that are able to support their own weight to a certain extent. That is, mechanical strength in the forming branches is able to withstand forces generated due to sedimentation flow as fluid is squeezed out from the holes in the network.

Note that doubling the attraction strength halves the slope of the boundary line,~\textit{i.e.}, the crossover P{\'e}clet number $\mathrm{Pe}_{\mathrm{c}} \propto \exp(C \epsilon \phi_{0})$ with $C$ a constant. This is indicative of attractions dominating the dynamics, or equivalently, of mechanical strength in the emerging network.

The linear trend is bounded from below by an effective percolation threshold $\phi_{\mathrm{p}} \approx 0.033$. It is clear that the system must be able to percolate to form a network structure and eventually a gel. However, below the green bounds, even a system with $F_{\mathrm{B}} = 0$ was not found to percolate. We obtained the respective $\epsilon$ boundaries by determining the minimum volume fraction required to observe percolation within $10^{3}t_{\mathrm{B}}$ in bulk for NH gels, using the methods described in Ref.~\cite{degraaf2019hydrodynamics}. Appendix~\ref{sec:percolation} provides the details of this analysis.

For all our considered $\mathrm{Pe}$ values, the system has equilibrated its gravitational profile in this time. It should be noted, however, that for $\mathrm{Pe} \downarrow 0$, at some point DLCA (purple curve in Fig.~\ref{fig:cross}) is expected to hold: we therefore did not extend the green lines fully to the left. Our percolation threshold depends slightly on the value of $\epsilon$, see Fig.~S6~\cite{supplement}, which controls cluster rearrangement. Larger $\epsilon$ compacts the forming clusters more and hinders in-cluster rearrangements, raising $\phi_{\mathrm{p}}$.

\begin{figure}[!htb]
\centering
\includegraphics[width=0.95\columnwidth]{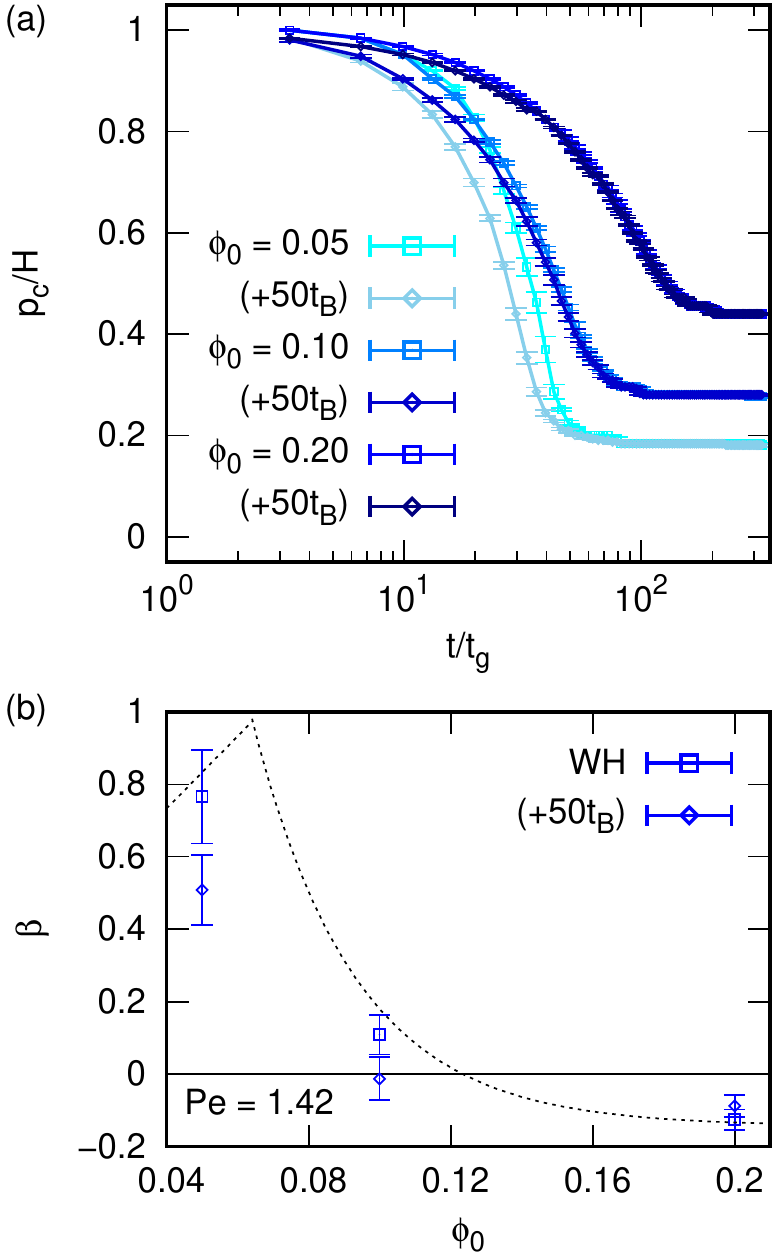}
\caption{\label{fig:preform}The effect of preforming the gel network by $50t_{\mathrm{B}}$ before switching on gravity. (a) Comparison of the interface $p_{\mathrm{c}}$ evolution (no-preforming, $\Square$, and preforming $\diamond$, respectively) as a function of time $t$ in terms of the gravitational time $t_{\mathrm{g}}$. The three initial volume fractions are as labelled for simulations with HIs, $\mathrm{Pe} = 1.42$, and $\epsilon = 10 k_{\mathrm{B}}T$. (b) Comparison of the associated exponent $\beta$ as a function of the initial volume fraction $\phi_{0}$. The dashed lines represent the fitted exponent and linear trend obtained from the full $\mathrm{Pe} = 1.42$ and $\epsilon = 10 k_{\mathrm{B}}T$ data without preforming, see Fig.~\ref{fig:exp_142}.}
\end{figure}

We tested the robustness of our result by preforming gels without gravity ($F_{B} = 0$) for $50 t_{\mathrm{B}}$ after equilibration. Subsequently switching on gravity, we found that the sedimentation is slightly more rapid at $\phi_{0} < \phi_{\mathrm{c}}$, see Fig.~\ref{fig:preform}a. We interpret this as being related to the fact that preforming leads to larger \textit{anisotropic} clusters, which have a higher rate of collective settling. Note that $\beta$ is slightly lowered for a given $\phi_{0}$, as the cluster growth should be diminished when the gel is allowed to preform. Nonetheless, the trend in $\beta$ is similar, also see Figs.~\ref{fig:crit}b and~\ref{fig:preform}. This similarity implies that $\phi_{\mathrm{c}}$ is a meaningful stability measure, though clearly the value of $\phi_{\mathrm{c}}$ will be shifted to lower values of $\phi_{0}$ by preforming the gel, as there will be greater mechanical strength at lower volume fractions.

\section{\label{sec:sediment}Properties of the Sediment}

As the nascent gels settle (rapidly or slowly), they become more compact, since both the mechanical strength and resistance to backflow increase. This eventually ($t \gg 100 t_{\mathrm{g}}$) leads to structures that are arrested ($v_{\mathrm{c}} \approx 0$) and can be characterized as colloidal gels. We do expect these gels to compact further at times longer than we can probe numerically.

\begin{figure}[!htb]
\centering
\includegraphics[width=0.95\columnwidth]{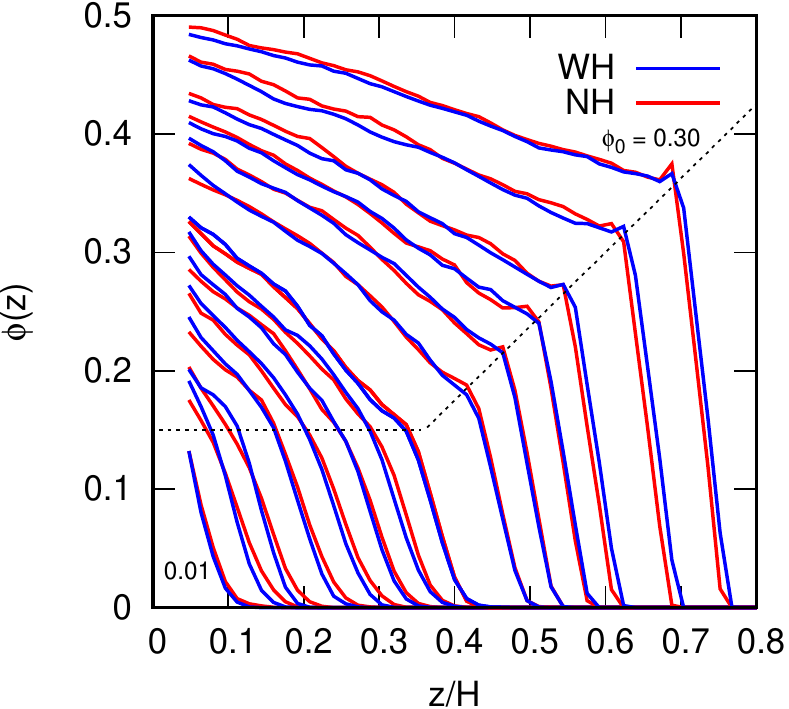}
\caption{\label{fig:profile}The average colloid concentration in a horizontal slice $\phi(z)$ as a function of the height $z$ (normalized by $H$) after sedimentation at $\mathrm{Pe} = 0.28$ and $\epsilon = 10 k_{\mathrm{B}}T$. The blue curves show the data with (WH) and the red curves without hydrodynamic interactions (NH). Results for $14$ initial volume fractions $\phi_{0}$ are provided, from left to right the values are $\phi_{0} = 0.01$, 0.018, 0.025, 0.038, 0.05, 0.063, 0.075, 0.088, 0.1, 0.125, 0.15, 0.175 0.2, 0.25, and 0.3. The two straight dashed lines are guides to the eyes, indicating a crossover in trend. The standard error is roughly three times the line width; it is not shown here to improve the presentation.}
\end{figure}

Figure~\ref{fig:profile} shows height profiles $\phi(z)$ in the gel state for our standard set of $\phi_{0}$ and $\mathrm{Pe} = 0.28$,~\textit{i.e.}, the typical reference point for our discussion thus far. At $\phi_{0} = 0.30$, $\phi(z)$ shows a narrow interfacial region at the top ($0.8 \gtrsim z \gtrsim 0.7$) where the density rises rapidly as $z$ decreases, followed by the main gel body in which the density increases slowly from $\phi \approx 0.4$ to 0.5. The presence of a density gradient recalls low-$\mathrm{Pe}$ results without HIs~\cite{whitmer2011sedimentation}. As can be appreciated from Appendix~\ref{sec:sedprof}, the result is robust with system size. However, Fig.~\ref{fig:profiles} in that appendix also reveals that the shape of $\phi(z)$ depends sensitively on the value of $\mathrm{Pe}$.

A clearly demarcated interfacial region is also evident in the $\phi(z)$ data at lower volume fractions. As $\phi_{0}$ decreases, the crossover from interface to gel body decreases linearly, and the density gradient in the latter becomes progressively less linear. When $\phi_{0}$ has dropped to 0.1, the crossover from interface to gel body has become a continuous `knee' rather than an abrupt change in slope. Interestingly, the cross-over point, in so far as it still can be discerned, becomes constant at $\phi_{0} < 0.1$. This feature is more pronounced for small values of $\mathrm{Pe}$, see Fig.~\ref{fig:profiles}a.

It is presently unclear what underlies the change in trend in $\phi(z)$ as a function of the initial volume fraction and of $\mathrm{Pe}$. However, it is clear that the explanation must be sought in terms of a mechanical description, as (surprisingly!) we find no systematic difference between gels obtained with and without HIs, despite the significant differences in settling rate between the two at low $\phi_{0}$.

\section{\label{sec:discussion}Connection with Experiment}

Lastly, let us turn to experiments on colloidal gels in narrow confinement~\cite{razali2017effects, tsurusawa2019direct}. To the best of our knowledge, only Razali~\textit{et~al.}~\cite{razali2017effects} systematically investigated the effect of buoyancy on the gelation and settling. They concluded that ``in small systems sedimentation is enhanced relative to non-gelling suspensions, although the rate of sedimentation is reduced when the strength of the attraction between the colloids is strong.'' Our results suggest that the enhanced sedimentation may be partially explained by backflow and cluster growth. The reduction in settling speed with increased attraction strength aligns with our finding that this shifts the stability region, thus slowing down the dynamics at fixed $\phi_{0} > \phi_{\mathrm{c}}$.

Turning to our sediments, see Figs.~\ref{fig:profile},~\ref{fig:profiles}, and~\ref{fig:profile_h}, we note that these are a long-lived metastable `gel' state. That is, there is an `open' network structure at intermediate volume fractions that supports its own weight against gravity for times considerably exceeding the gravitational and Brownian time. We note that for the realistic experimental values of $\mathrm{Pe} = 0.28$~\cite{zhou2018onset} and $\mathrm{Pe} = 1.41$~\cite{harich2016gravitational}, we have a significant density gradient in our `gel'. By contrast, the \textit{initial} density profile of experimentally reported colloidal gels is typically homogeneous~\cite{secchi2014time, harich2016gravitational, zhou2018onset}. However, it should be noted that those measurements are carried out over much larger length scales. It may be that our `gels' are too small to exhibit a constant density regime, even taking into account the substantially taller simulation volumes reported in Appendix~\ref{sec:sedprof}. The discrepancy between experiment and simulation may also be due to the effect of frictional constraints on interparticle motion~\cite{wang2019surface, nguyen2020computer, immink2020using} or aggregation~\cite{turetta2022role}. The source of the mismatch makes an interesting subject for future study. However, here, we have chosen not to analyze the origin of the shape of the $\phi(z)$ profiles further, as it does not capture the experiments of interest~\cite{harich2016gravitational}.

\section{\label{sec:outro}Conclusion and Outlook}

Summarizing, it is clear that \textit{short-time} buoyancy-induced flows can significantly impact colloidal gelation. We have identified a dynamic stability criterion for the colloid volume fraction at which the system gels, which depends on attraction strength and the ratio of buoyant settling versus diffusion. The stability crossover emerges as a competition between cluster growth, leading to a low-volume fraction acceleration of settling, and intermediate volume-fraction network formation. The latter imparts mechanical strength to the nascent gel, that causes it to experience hydrodynamic drag as a porous material rather than a collection of disconnected clusters.

However, interestingly, none of the drastic short-time dynamics induced by backflows strongly impact the settled transient gel structure. Future work that includes (lubricated) frictional interparticle interactions and tackles the computational challenges of simulating larger system sizes, may bring better agreement with experimental observations. This is a necessary first step toward predicting and subsequently improving the shelf life of colloidal gel products.


\begin{acknowledgments}
We thank EPSRC for funding through Programme Grant (EP/J007404/1) and the International Fine Particles Research Institute (WCKP) for support; JdG further acknowledges the NWO for funding through Grant No.~OCENW.KLEIN.354, as well as the EU through a Marie Sk{\l}odowska-Curie Intra European Fellowship (G.A.~No.~654916) within Horizon 2020. We are grateful to S.~Bindgen, E.~Koos, P.~Royall, J.~Swan, and X.~Zhou for useful discussions. 

Author contributions: Conceptualization, J.d.G., W.C.K.P., \& M.H.; Methodology, J.d.G. \& M.H.; Software, J.d.G. \& M.H.; Validation, J.d.G.; Investigation, J.d.G (lead), M.H. \& K.W.T. (supporting); Writing -- Original Draft, J.d.G. \& M.H.; Writing -- Review \& Editing, W.C.K.P.; Funding Acquisition, J.d.G. \& W.C.K.P.; Resources, J.d.G. \& W.C.K.P.; Supervision, J.d.G. 

An open-data package containing the means to reproduce the figures and overall results of the simulations is available at: [10.24416/UU01-Z9XKDO].
\end{acknowledgments}



\appendix

\section{\label{sec:videos}Description of Accompanying Videos}

The two videos accompanying Fig.~1 from the main text show the evolution of the colloidal suspension in a thin (2$\sigma$) slice though the center of the simulation box with the colloids indicated in red (NH; \texttt{collapse\_lv.mp4}) or blue (WH; \texttt{collapse\_lb.mp4}); $\mathrm{Pe} = 0.28$ and $\epsilon = 10 k_{\mathrm{B}}T$. In both movies the initial volume fraction of the sample increases from left to right: $\phi_{0} = 0.025$, $0.05$, $0.075$, $0.1$, $0.15$, and $0.3$, respectively. The frame rate is such that for each second $10 t_{\mathrm{g}}$ elapses, with $t_{\mathrm{g}}$ the gravitational time (see main text).

The video \texttt{compare\_preform.mp4} shows the sedimentation for the colloidal suspension WH in a similar representation for $\phi_{0} = 0.1$, $\mathrm{Pe} = 1.42$, and $\epsilon = 10 k_{\mathrm{B}}T$. The difference between the blue representation (left) and the cyan one (right) is that the latter is preformed for $50t_{\mathrm{B}}$ with $t_{\mathrm{B}}$ the Brownian time. It should be noted that the interface of the preformed gel sediments faster.

\section{\label{sec:scale}Scale-Independent Settling Dynamics}

\begin{figure}[!htb]
\centering
\includegraphics[width=0.95\columnwidth]{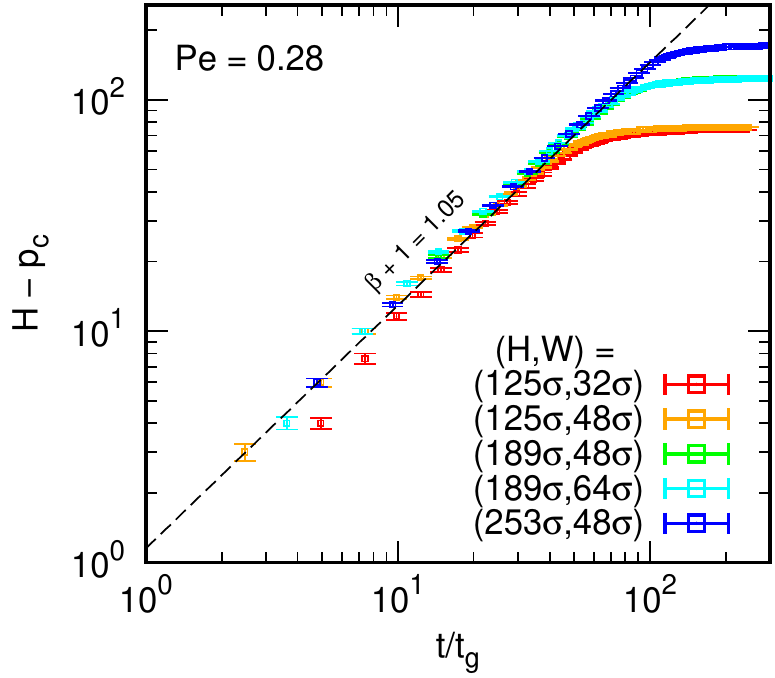}
\caption{\label{fig:fss}The effect of system size on the initial power-law behavior of the interface between the colloid-rich and colloid-poor region for $\mathrm{Pe} = 0.28$, $\phi_{0} = 0.1$, and $\epsilon = 10 k_{\mathrm{B}}T$. The height of the colloid-poor region $(H - p_{\mathrm{c}})$ is given as a function of the reduced time $t/t_{\mathrm{g}}$ for five different box sizes, as labelled. The dashed lines indicates a power-law fit that works well for all box sizes, from which we find $\beta \approx 0.05$. All simulations in this figure were performed with HIs.}
\end{figure}

Figure~\ref{fig:fss} reveals that the box size in which we performed the simulations did not measurably impact the settling coefficient $\beta$. Here, we specifically focused on the initial density $\phi_{0} = 0.1$ for $\mathrm{Pe} = 0.28$ and $\epsilon = 10 k_{\mathrm{B}}T$, because this is where we locate $\phi_{\mathrm{c}}$ ($\beta \approx 0$). If there was a significant effect of the system size/shape on the critical volume fraction, it would have been revealed here. We note that there is a slightly better defined initial power-law behavior for the larger systems. From the cumulative results we obtain a value of $\beta \approx 0.05$, which is sufficiently close to our original measurement.

In addition to the analysis performed using simulations accounting for HIs and flow, we performed regular Langevin dynamics simulations for two additional box sizes: effective heights $H = 125\sigma$ and $H = 253\sigma$, respectively, with a square base of length $48\sigma$ in both cases. The effect on the $\beta$ parameter that follows from these is minimal, as can be appreciated from the inset to Fig.~\ref{fig:exp_142}. This is unsurprising, any effects would be most strongly expressed in simulations with HIs. We mention the result here for completeness. The goal of these simulations was to chart the effect of box height $H$ on the heigh profile $\phi(z)$ in the long-time sediment. The profiles we obtained were qualitatively similar, as shown in Appendix~\ref{sec:sedprof}.

\section{\label{sec:exponents}Additional Sedimentation Exponents}

\begin{figure}[!htb]
\centering
\includegraphics[width=0.95\columnwidth]{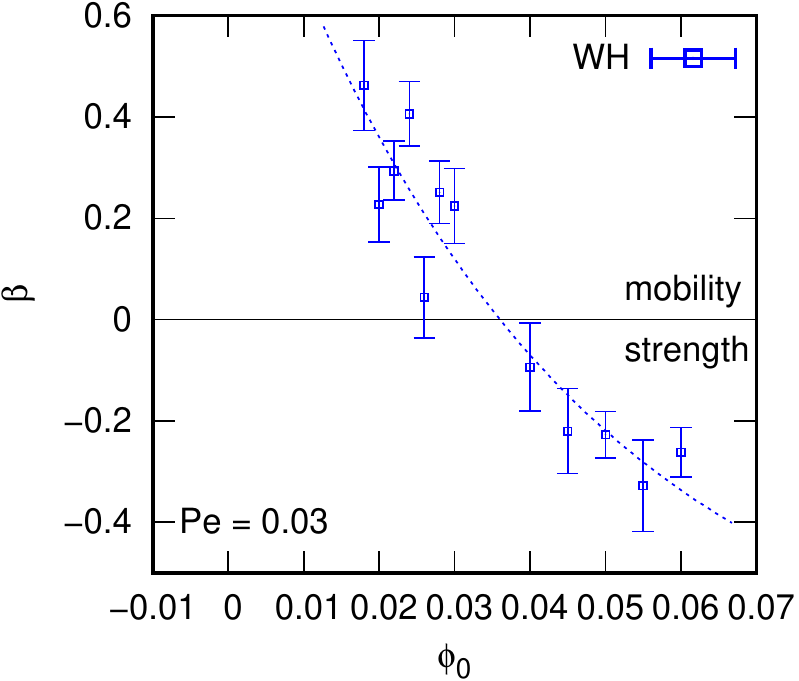}
\caption{\label{fig:exp_003}The exponent $\beta$ as a function of $\phi_{0}$ for $\mathrm{Pe} = 0.03$. Blue squares indicate data obtained with hydrodynamics (WH). Due to the long run-time of the simulations, the non-hydrodynamic (least insightful) curves were not computed for this value of $\mathrm{Pe}$. The error bars provide the standard error and the dashed curve is an exponential fit that additionally serves as a guide to the eyes.}
\end{figure}

\begin{figure}[!htb]
\centering
\includegraphics[width=0.95\columnwidth]{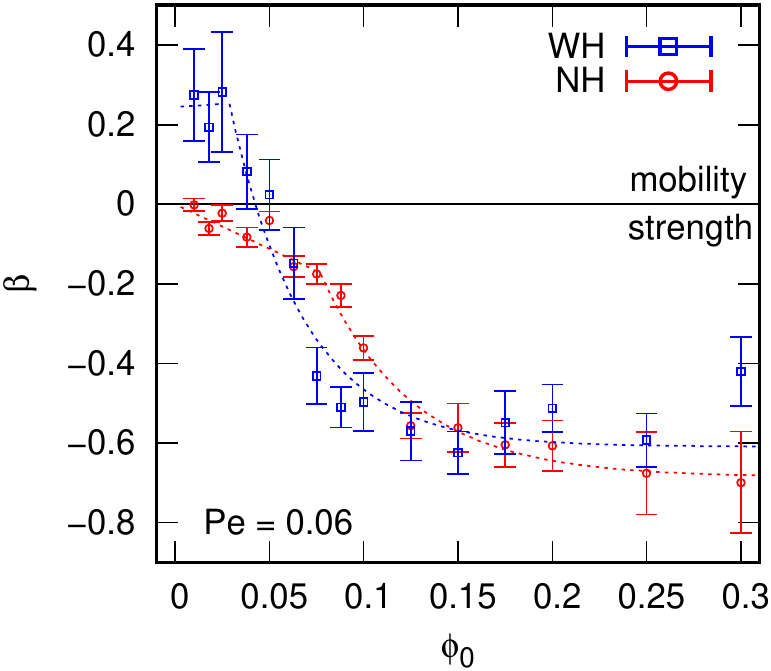}
\caption{\label{fig:exp_006}The exponent $\beta$ as a function of $\phi_{0}$ for $\mathrm{Pe} = 0.06$. Blue squares indicate data obtained with hydrodynamics (WH) and red circles the non-hydrodynamic (NH) data. The error bars provide the standard error and the dashed curves are piece-wise fits of a linear and exponential function.}
\end{figure}

\begin{figure}[!htb]
\centering
\includegraphics[width=0.95\columnwidth]{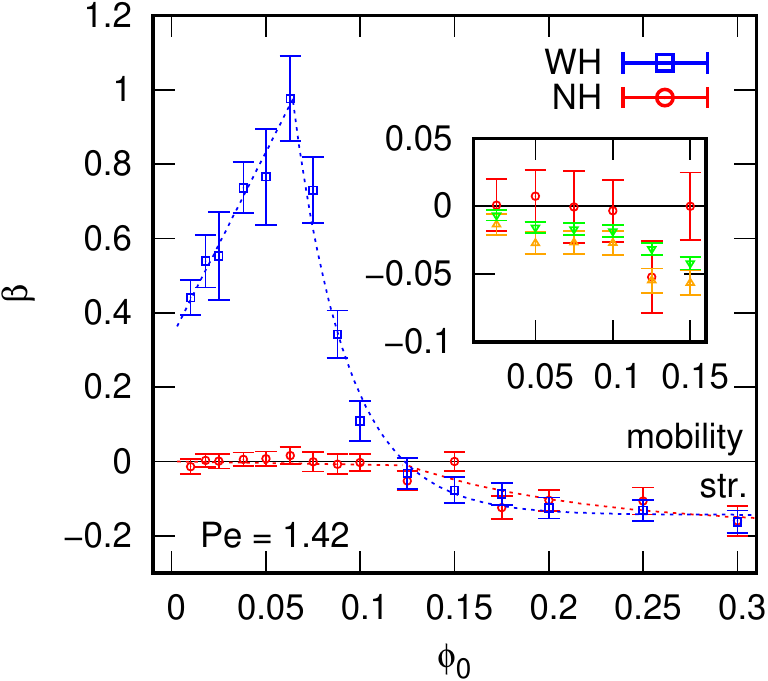}
\caption{\label{fig:exp_142}The exponent $\beta$ as a function of $\phi_{0}$ for $\mathrm{Pe} = 1.42$ in a box of height $H = 125\sigma$ and width of $32\sigma$. Blue squares indicate data obtained with hydrodynamics (WH) and red circles the non-hydrodynamic (NH) data. The error bars provide the standard error and the dashed curves are piece-wise fits of a linear and exponential function. The inset shows two additional NH simulation results for effective heights $H = 125\sigma$ (orange, \textcolor{orange}{$\triangle$}) and $H = 253\sigma$ (green, \textcolor{green}{$\triangledown$}), respectively, with a square base of length $48\sigma$.}
\end{figure}

In this appendix, we provide additional information on the exponent $\beta$ that describes the power-law dependence of the initial settling velocity. Here, we present this as a function of $\phi_{0}$ for $\mathrm{Pe} = 0.03$, $0.06$, and $1.42$ in Figs.~\ref{fig:exp_003},~\ref{fig:exp_006}, and~\ref{fig:exp_142}, respectively, where $\mathrm{Pe}$ indicates the gravitational P{\'e}clet number.

Only the $\epsilon = 10 k_{\mathrm{B}}T$ result is shown in Fig.~\ref{fig:exp_003}, the result for $\epsilon = 20 k_{\mathrm{B}}T$ is analogous. Accelerated sedimentation is found for all values of $\mathrm{Pe}$ in systems with HIs. The guides to the eye are a composite between a fitted linear function for small $\phi_{0}$ and an exponential decay, respectively. Both fits were obtained using a least-squares approach. Note the pronounced linear trend of the low-$\phi_{0}$ data for $\mathrm{Pe} = 1.42$ in Fig.~\ref{fig:exp_142}, which is referenced in the main text.

\section{\label{sec:percolation}Percolation Measurements}

\begin{figure}[!htb]
\centering
\includegraphics[width=0.95\columnwidth]{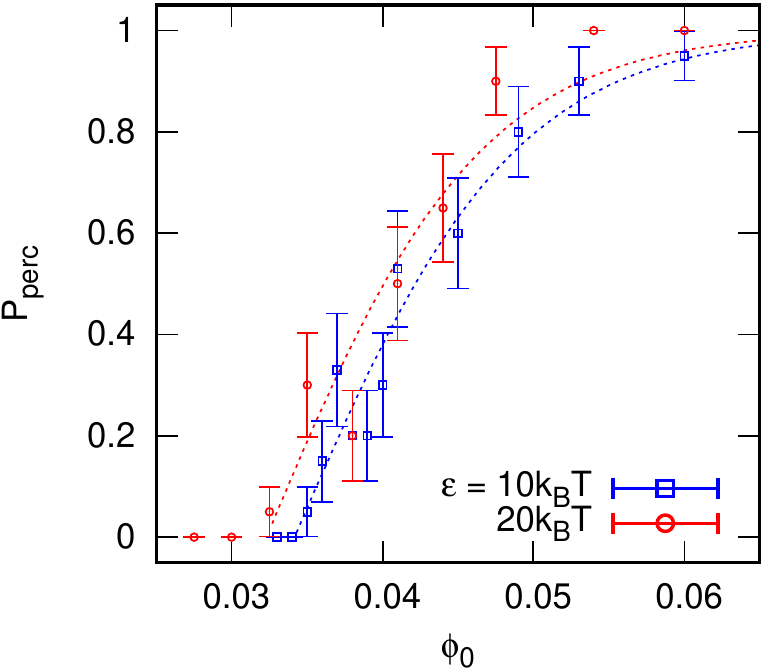}
\caption{\label{fig:percolation}Estimation of the percolation threshold. The probability of percolating $P_{\mathrm{perc}}$ as a function of the initial volume fraction $\phi_{0}$ for two interaction strengths: $\epsilon = 10 k_{\mathrm{B}}T$ (blue) and $\epsilon = 20 k_{\mathrm{B}}T$ (red). Error bars indicate the standard error of the mean and the dashed lines are fits to the data with the function $\tanh[ A ( \phi_{0} - B )]$ where $A$ and $B$ are constants.}
\end{figure}

Figure~\ref{fig:percolation} shows the result of our analysis of the percolation threshold, on which we base our vertical gelation boundaries in Fig.~3 of the main text. To obtain this quantity, we simulated 1,000 colloids in a cubic simulation volume for several edge lengths, such that we obtained the desired values of $\phi_{0}$ shown in the figure. The systems were prepared in the usual manner (main text and Ref.~\cite{degraaf2019hydrodynamics}) and were allowed to gel for the somewhat arbitrary, but very long time of $10^{3} t_{\mathrm{B}}$. At this point, we determined whether the system had percolated, which we define by asserting that a cluster exists that self-connects in one direction across the periodic simulation volume. We use an inter-particle spacing of $1.05\sigma$ to determine whether a particle is part of a cluster or not, with $\sigma$ the diameter in our high-exponential Lennard-Jones potential (see main text). This measurement was repeated 20 times for each point in our data set, from which we obtained our error bars, as shown in Fig.~\ref{fig:percolation}.

The data revealed a lower-bound to $\phi_{0}$ beyond which we did not find any instances of systems that had percolated. Clearly, our data shows finite-size and finite-time effects, as there is a sizeable transition zone where the system has a finite, non-unity probability to percolate. In principle, the percolation analysis can be improved by scaling out the finite system size, which should reveal a sharper transition. However, we chose not to do so here, as obtaining this level of data is a costly process and further improving it does not serve the purpose of our work. We fitted the data using $\tanh[ A ( \phi_{0} - B )]$, where $A$ and $B$ are constants, to determine the point where the system did not percolate. The shape appeared to capture our trend well, but is otherwise not physically motivated. The fitting gave the following transition points $\phi_{\mathrm{p}} = 0.034 \pm 0.001$ and $\phi_{\mathrm{p}} = 0.032 \pm 0.001$ for $\epsilon = 10 k_{\mathrm{B}}T$ and $\epsilon = 20 k_{\mathrm{B}}T$, respectively, see main text.

Note that surprisingly, the $\epsilon = 20 k_{\mathrm{B}}T$ system has a slightly lower percolation threshold than the $\epsilon = 10 k_{\mathrm{B}}T$ system. We suspect that the higher interaction strength leads to slower rearrangements of the forming gel strands. As a consequence, the clusters that form remain more extended, allowing them to percolate more readily. However, the effect is subtle since the shift in the average trend is less than 10\% of the signal.

\section{\label{sec:sedprof}Additional Sedimentation Profiles}

\begin{figure}[!htb]
\centering
\includegraphics[width=0.95\columnwidth]{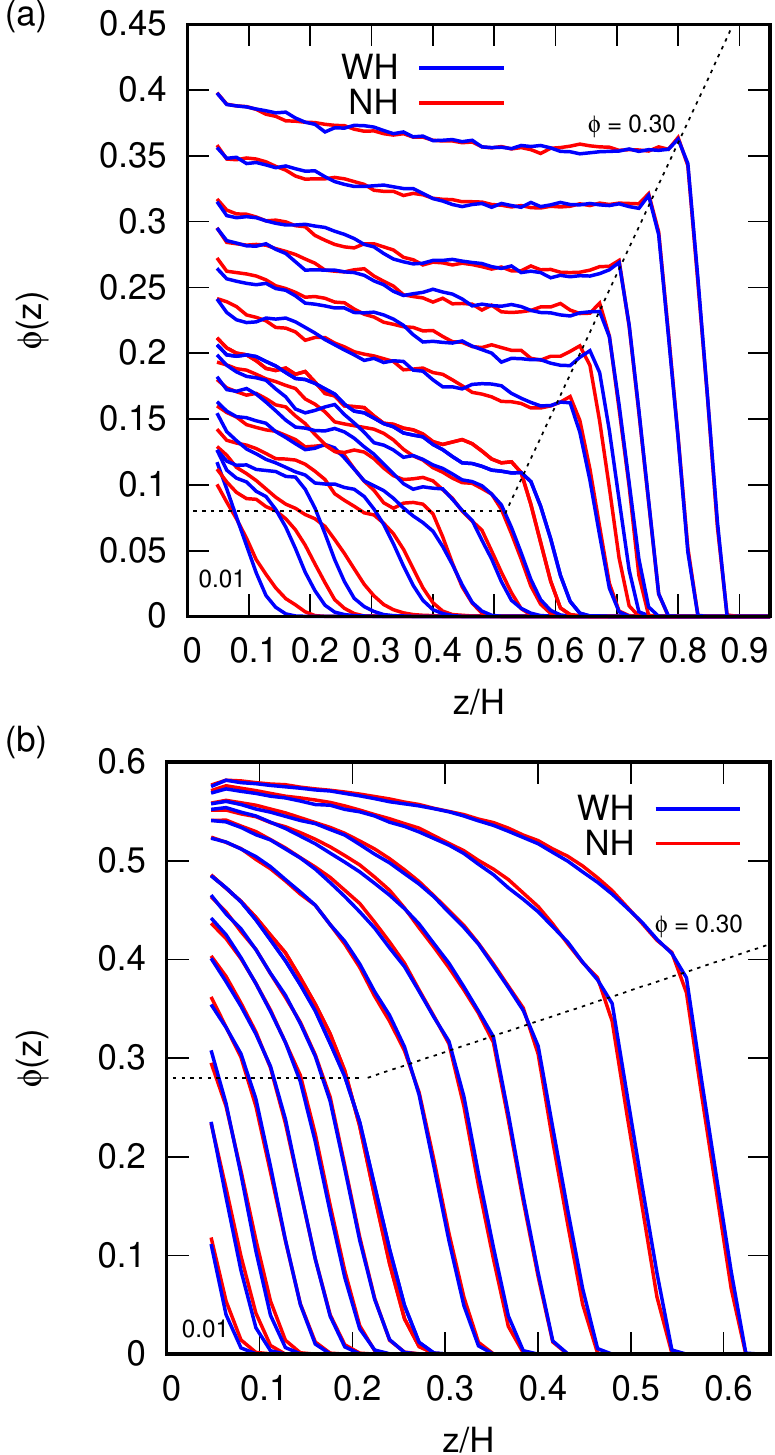}
\caption{\label{fig:profiles}The average colloid concentration in a horizontal slice $\phi(z)$ as a function of the height $z$ (normalized by $H$) after sedimentation at (a) $\mathrm{Pe} = 0.03$ and (b) $\mathrm{Pe} = 1.42$, respectively; in both cases $\epsilon = 10 k_{\mathrm{B}}T$. The blue curves show the data with (WH) and the red curves without hydrodynamic interactions (NH). Results for $14$ initial volume fractions $\phi_{0}$ are provided, from left to right the values are $\phi_{0} = 0.01$, 0.018, 0.025, 0.038, 0.05, 0.063, 0.075, 0.088, 0.1, 0.125, 0.15, 0.175 0.2, 0.25, and 0.3. The two straight dashed lines are guides to the eyes, indicating a crossover in trend. The standard error it is not shown here to improve the presentation, but comparable to the fluctuations on the trends.}
\end{figure}

Figure~\ref{fig:profiles} show a crossover in the shape of the sedimented suspension that mirrors that shown in Fig.~\ref{fig:profile} for values of the P{\'e}clet number of $\mathrm{Pe} = 0.03$ and $\mathrm{Pe} = 1.42$, respectively. This crossover is located at $\phi_{0} \approx 0.08$.

Note that depending on the value of $\mathrm{Pe}$, the shape of the sediment below the meniscus --- to the left of the sharp increase in density --- differs substantially. In Fig.~\ref{fig:profile} of the main text, the profile is clearly linear, while in Fig.~\ref{fig:profiles}a there is a concave quality to it and Fig.~\ref{fig:profiles}b presents a convex shape. To give context to this, a linear density profile is similar to the density trend found for an initially homogenous, linearly elastic solid under gravitational compression. The flattening off of the profile for low values of the P{\'e}clet number indicates that above a certain volume fraction, for low gravitational strength, the gel can support (some of) its own weight without becoming strongly compressed. Conversely at high $\mathrm{Pe}$, compaction at low values of $z$ is needed to support the weight that is rests on top of the network.

\begin{figure}[!h]
\centering
\includegraphics[width=0.95\columnwidth]{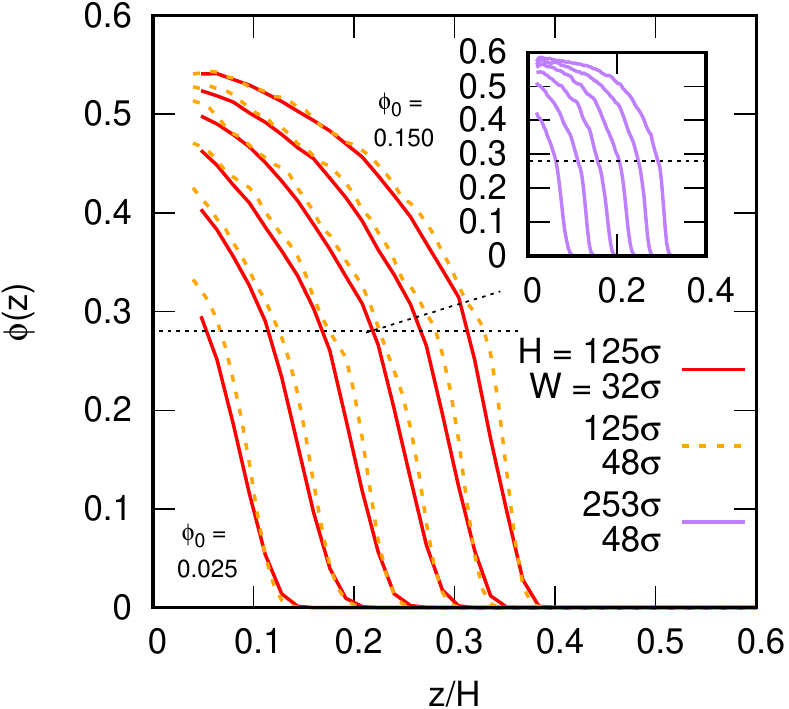}
\caption{\label{fig:profile_h}The average colloid concentration profile $\phi(z)$ as a function of the reduced height $z/H$ in the sediments that form at $\mathrm{Pe} = 1.42$ and $\epsilon = 10 k_{\mathrm{B}}T$. The red data in the main panel is for our standard box dimensions of $H = 125\sigma$ and basal length $32\sigma$, while the orange dashed curve shows the result for a basal width of $48\sigma$ and the same height $H$. The inset shows data (purple curves) for an effective height of $H = 253\sigma$ and a basal width of $48\sigma$. All data was obtained without accounting for HIs. Results for $5$ initial volume fractions $\phi_{0}$ are provided, from left to right the values are $\phi_{0} = 0.025$, 0.050, 0.075, 0.100, 0.125, and 0.150. The two straight dashed lines are guides to the eyes, indicating a crossover in trends. The standard error it is not shown here to improve the presentation, but comparable to the fluctuations on the trends.}
\end{figure}

Finally, we consider the effect of box dimensions on the sediment in Fig.~\ref{fig:profile_h}. Here, we used the highest gravitational strength and volume fractions up to $\phi_{0} = 0.15$ in simulations without HIs to have the gel sediments form in a reasonable time. We note from the main panel that widening the box has little effect on the shape of the sediment. However, within the error, the upward turn in `knee' in $\phi(z)$ that indicates the crossover between the gel body and the colloid-dense and colloid-poor interface is less pronounced; we have extended the horizontal guide to the eye to indicate this. The shape of the profile for a box that is nearly double the height ($H = 253\sigma$) is similar, but the presence of a knee is even less pronounced. We conclude that --- at least for the simulations that we have checked in this manner --- the system size does not appreciably impact the qualitative trend in observed for the $\phi(z)$ profile of the sediment.

\end{document}